\documentclass[12pt]{article}

\usepackage{amssymb}
\usepackage{amsmath}
\usepackage{latexsym}

\catcode`@=11
\renewcommand{\section}{\@startsection{section}{1}{0pt}{\medskipamount}
{\medskipamount}{\large\bf}}
\numberwithin{equation}{section}
\catcode`@=12

\def\beq{\begin{eqnarray}}    
\def\eeq{\end{eqnarray}}      

\def\ln{\,\mbox{ln}\,}                  

\def\pa{\partial}                       

\def\={\ =\ }

\begin{document}

\begin{center}

{\Large\bf Extended sigma-model in nontrivially deformed field-antifield formalism}

\vspace{18mm}

{\Large Igor A. Batalin$^{(a,b)}\footnote{E-mail:
batalin@lpi.ru}$\;,
Peter M. Lavrov$^{(b, c)}\footnote{E-mail:
lavrov@tspu.edu.ru}$\;
}

\vspace{8mm}

\noindent ${{}^{(a)}}$
{\em P.N. Lebedev Physical Institute,\\
Leninsky Prospect \ 53, 119 991 Moscow, Russia}

\noindent  ${{}^{(b)}}
${\em
Tomsk State Pedagogical University,\\
Kievskaya St.\ 60, 634061 Tomsk, Russia}

\noindent  ${{}^{(c)}}
${\em
National Research Tomsk State  University,\\
Lenin Av.\ 36, 634050 Tomsk, Russia}

\vspace{20mm}

\begin{abstract}
\noindent
We  propose  an action  for  the  extended  sigma  - models  in  the  most
general   setting of  the  kinetic  term  allowed  in  the  nontrivially  deformed   field -
antifield  formalism. We  show  that  the  classical  motion equations  do  naturally
take  their
desired  canonical form.
\end{abstract}

\end{center}

\vfill

\\

\newpage

\section{Introduction and summary}

The  field-antifield  formalism  \cite{BV,BV1}  is  known  as  the most  powerful
method for covariant quantization  of  general  gauge  theories.
The  method  operates   with two  fundamental
objects:   the   antibracket   and   the  odd  nilpotent   second-order
Delta  operator.  In the standard  version  of  the  method,
it  is  formulated  in  the Darboux coordinates.  Various
local  deformations  of  the  Delta  operator \cite{BT}  and  antibracket
\cite{LSh,KoT,KoT1,KoT2}  elucidate the geometry underlying
the   field-antifield   formalism  \cite{AKSZ}.  The  deformed
Delta  operator and  antibracket   do  play  the  central  role  in  formulating  the  path
integral  version  of the  nontrivially deformed  field-antifield   formalism \cite{BB2}.
It has  been shown  recently \cite{BL} that  the nontrivially deformed  formalism
does exist,  as a consistent approach,  at  the classical level.
In the present article, we  confirm  the  latter result as applied
to the important case of the  topological sigma-models
\cite{AKSZ,Witten,SS,CF,LO,BM4,BM5,BLN,BG}.
We  propose  an  action of  the  deformed sigma-models  in the most general
setting   for  the kinetic  term  allowed in  the  formalism,
and show that the motion equations do naturally take their desired canonical
form.
We define the kinetic part of the action directly in terms of the zero modes
for the $\tau$- extended Euler operator  $N_{\tau}$, that generalizes naturally
the standard power counting operator.  We show
that  these zero  modes, although they are nontrivial functions of the
original phase variables, do
satisfy very simple  antibracket  algebra,  so that they  generate
effectively their own antisymplectic
structure, very similar to the original one. The latter  circumstance  makes
it possible to deduce the
desired canonical form of the motion equations in a very simple and  natural
way.  Moreover, we have
deduced  the so - called canonical  "Ward Identities"  the complete action
does satisfy to  at the general
(arbitrary) trajectory in the antisymplectic  phase space. These identities,
in  their own  turn, are  necessary
when  deducing  the functional master equation for  the complete action .
Thus,  we  have  got  to a
considerable new  progress in our study of the canonical structure in
topological  field theories.
\\

\section{Elements of the nontrivially deformed field-antifield formalism}

Let $\{ Z^{A}, A = 1,...,2 M \}$, \;$\varepsilon( Z^{A} ) = \varepsilon_{A}$, be
original field-antifield phase variables  with
constant invertible antisymplectic metric , $E_{AB}, \;\varepsilon( E_{AB} ) =
\varepsilon_{A} + \varepsilon_{B} + 1$,
\beq
\label{E}
E_{AB} = -  E_{BA} (-1)^{ \varepsilon_{A} \varepsilon_{B} }  =  {\rm const}( Z ). 
\eeq
Thereby, we have defined the antibracket,
\beq
\label{AnB}
( F, G )  =  F \overleftarrow{\pa}_{A} E^{AB} \overrightarrow{\pa}_{B} G,\quad
E^{AB}E_{BC}  =  \delta^{A}_{C},    
\eeq
\beq
\varepsilon(E^{AB})  =  \varepsilon_{A} + \varepsilon_{B} + 1, \quad  E^{AB}  =
- E^{BA} (-1)^{ ( \varepsilon_A + 1) ( \varepsilon_{B} + 1 ) },    
\eeq
with  all  standard  properties.  Then, the Fermionic  Delta - operator,
$\Delta,  \varepsilon( \Delta ) =  1$, is given by
\beq
\label{Delt}
\Delta = \frac{1}{2} (-1)^{ \varepsilon_{A} } \pa_{A} E^{AB} \pa_{B},  \quad
\Delta^{2} = 0.   
\eeq
Vice  versa, the  antibracket  (\ref{AnB})  can  be deduced
from  the  Delta - operator (\ref{Delt}), \beq \label{AnB2}
( F, G )  =  (-1)^{ \varepsilon_{F} } [ [ \Delta, F ], G ] \cdot 1. 
\eeq

Now,  let  $\{ t,  \theta \}$,  $\varepsilon( t ) = 0,\; \varepsilon( \theta ) = 1$,
be a pair of new  antisymplectic  variables.
Let $( F, G )_{\tau}$  and  $\Delta_{\tau}$  be  the  extended  counterparts  to
$( F, G )$  and  $\Delta$,  respectively,
\beq
\label{AnBtau}
( F, G )_{\tau}  =  t^{2} ( F, G )  +  ( N_{\tau} F ) \pa_{\theta} G  -
F\overleftarrow{\pa}_{\theta} N_{\tau} G,     
\eeq
\beq
\label{Delttau}
\Delta_{\tau}  =  t^{2} \Delta  +  N_{\tau} \pa_{\theta}, \quad
N_{\tau}  =  N  + t  \pa_{t},  \quad N = N^{A} \pa_{A},    
\eeq
\beq
\label{e2.8}
\Delta_{\tau}^{2}  =  0,   \quad   [ \Delta, N ] = 2 \Delta, \quad      [
\Delta_{\tau}, N_{\tau} ] = 0.    
\eeq
Vice versa, the extended antibracket  $( F, G )_{\tau}$  can be deduced from
the extended Delta-operator, $\Delta_{\tau}$,
\beq
( F, G )_{\tau}=(-1)^{ \varepsilon_{F} } [ [ \Delta_{\tau}, F], G ] \cdot 1.    
\eeq

Now, let the antisymplectic phase variables  $\{ Z^{A} ; t, \theta \}$ be
functions of  $2 n$ Bosonic variables  $u^{a}$
and $2 n$ Fermionic variables $\xi^{a}$,
\beq
Z^{A}  =  Z^{A}( u, \xi ),  \quad    t  =  t( u, \xi ),  \quad   \theta  =  \theta( u,
\xi ).      
\eeq
Let   $D$,  $\varepsilon( D )  =  1$,  be  the  De  Rham   differential   in
the  variables  $u, \xi$,
\beq
\label{D}
D  =  \xi^{a}\pa_{a},  \quad   \pa_{a} = \frac{\pa}{ \pa u^{a}}, \quad  D^{2}  =  0.
\eeq
Further, let  $\kappa$  be a  Bosonic  deformation  parameter,  and
$\Delta_{\tau *}$  be  the extended  trivially deformed  Delta  operator
\beq
\Delta_{\tau *}  = \Delta_{\tau} ( 1 - \kappa N_{\tau} )^{-1} ,  \quad
\Delta_{\tau * }^{2}  =  0,    
\eeq
and let $T$ be the corresponding trivial deformation operator,
\beq
T = 1 + \kappa \theta \Delta_{\tau *},   \quad     T^{-1} = 1 -  \kappa \theta
\Delta_{\tau},      
\eeq
so that the trivially deformed extended antibracket  and  Delta  operator
rewrite as
\beq
( F, G )_{\tau *}  =  T^{-1} ( T F , T G )_{\tau}  =  ( F, G )_{\tau}  + (
\kappa  N_{\tau} F ) ( \Delta_{\tau *} G ) +
( \Delta_{\tau *} F ) ( \kappa  N_{\tau} G ) (-1)^{ \varepsilon_{F} },  
\eeq
\beq
\Delta_{\tau *}   =   T^{-1}  \Delta_{\tau}  T.      
\eeq

\section{Trivially deformed sigma-model}

The  extended  trivially  deformed  sigma - model  is  formulated via  the
action
\beq
\label{S}
\Sigma  =  \int  [ du ] [ d\xi ] \mathcal{ L },       
\eeq
where the  Lagrangian  $\mathcal{ L }$  is  defined  via  $\mathcal{ S }$
satisfying   the  extended trivially deformed master equation,
\beq
\label{Stau*}
( \mathcal{ S }, \mathcal{ S } )_{\tau *} = 0,    
\eeq
or equivalently,
\beq
\label{Stau}
( T \mathcal{ S },  T \mathcal{ S } )_{\tau} = 0.   
\eeq

One should seek for a solution to  the equation (\ref{Stau*})/(\ref{Stau}) in the form
\beq
\mathcal{ S }  = \sum_{k = -2}^{\infty} \mathcal{ S }_{(k|0)} t^{k} +
 \theta \sum_{k = 1}^{\infty} \mathcal{S}_{(k|1)} t^{k},     
\eeq
with  the  component $ \mathcal{ S }_{ (-2|0) }  =  \mathcal{ S }^*$
identified  with  the
nontrivially  deformed  $\mathcal{ S }^*$,
\beq
( \mathcal{ S }^*, \mathcal{ S }^* )_{*}  =  0,      
\eeq
where  the  nontrivially deformed antibracket  is  given  by
\beq
( F, G )_{*}  =  ( F, G )  +  ( \kappa  (N - 2) F ) ( \Delta _{*} G) )  +
( \Delta_{*} F ) ( \kappa  (N - 2 ) G ) (-1)^{ \varepsilon_{F} }. 
\eeq
\beq
\Delta_{*}  =  \Delta ( 1 - \kappa ( N - 2 ) )^{-1}.     
\eeq

Let  us consider our  new Lagrangian  for  the  deformed  sigma-model  in
the  most  general  setting,
\beq
\label{e1}
\mathcal{L}  = \frac{1}{2} \bar{Z}^{A} E_{AB} D \bar{Z}^{B} (-1)^{
\varepsilon_{B} }  +
\frac{1}{2} ( \theta D \ln t + \ln t D \theta )  +  T \mathcal{ S },  
\eeq
where
\beq \label{e2}
\bar{Z}^{A}  = \exp\{ - ( \ln t ) N \} Z^{A},  \quad  N_{\tau} \bar{Z}^{A}=0,    
\eeq
is the zero mode for  the $N_{\tau}$ operator.   By making a variation $\delta Z^{C}$
in (\ref{e1}), we  find
\beq
\label{e3}
( \pa_{C} \bar{Z}^{A} ) \;E_{AB} \;D \bar{Z}^{B} (-1)^{ \varepsilon_{B} }  +
\pa_{C} ( T \mathcal{ S } )  =  0.   
\eeq
By  multiplying  the  equation  (\ref{e3})  by  the coefficients   $N^{C}$  from  the
left, we  get
\beq
\label{e4}
- \left( t\pa_t{\bar Z}^A\right)\; E_{AB}\; D \bar{Z}^{B} (-1)^{ \varepsilon_{B} }
+  N ( T \mathcal{ S } )  =  0.     
\eeq
On  the  other  hand,  by  making  a  variation  $\delta \ln t$   in   (\ref{e1}),  we
have
\beq
\label{e5}
\left(t\pa_t {\bar Z}^A\right)\; E_{AB}\; D \bar{Z}^{B} (-1)^{ \varepsilon_{B} }  +
t\pa_t (T \mathcal{ S })   +  D \theta   =  0,      
\eeq
It  follows  from  (\ref{e4}),  (\ref{e5})  that
\beq
\label{e6}
D \theta + N_{\tau} (T \mathcal{ S } )  =  0,     
\eeq
or,  equivalently,  in terms  of  the  extended   $\tau$ - antibracket,
\beq
\label{e7}
D \theta + ( T \mathcal{ S }, \theta )_{\tau}  =  0.     
\eeq
That  is  exactly  the  desired  canonical  motion  equation  for  the
variable  $\theta$.   The  two  other  canonical
motion  equations  do follow  from  (\ref{e1})  in  the usual   way,  as  well,
\beq
\label{e8}
D \ln t  + \pa_{\theta} ( T \mathcal{ S } )  =  0,     
\eeq
or in its canonical form,
\beq
\label{e9}
D \ln t + ( T \mathcal{ S }, \ln t )_{\tau}  =  0,        
\eeq
together with the canonical form of the equation  (\ref{e3}),
\beq
\label{e10}
D Z^{A} + ( T \mathcal{ S }, Z^{A} )_{\tau}  =  0.         
\eeq
The complete set of the canonical equations (\ref{e7}), (\ref{e9}), (\ref{e10})
tells us that the action (\ref{e1})  yields the
correct  general  canonical description  to  the  deformed  sigma - model
at  the  classical  level.

For the sake of completeness, an  explicit derivation of the
canonical form (\ref{e10}) directly from the original one (\ref{e3})
is given below. As the operator $(N_{\tau}-2)$ does differentiate
the antibracket (due to the second in (\ref{e2.8})),  we have \beq
\label{e3.18} ( N_{\tau} + 2 ) ( {\bar Z}^{A}, {\bar Z}^{B} )  =  0,
\quad \left.( {\bar Z}^{A},
\bar{Z}^{B} )\right |_{t = 1}  =  E^{AB}.      
\eeq
It follows from (\ref{e3.18}) that,  within the class of
regular functions of $\ln t$,
\beq
\label{e3.19}
( {\bar Z}^{A},{\bar Z}^{B} ) =
t^{-2}  \exp\{- ( \ln t ) N \}  E^{AB}   = t^{-2} E^{AB}.     
\eeq
The latter rewrites in the explicit form, \beq \label{e3.20}
{\bar Z}^{A} \overleftarrow{\pa}_{C} E^{CD} \overrightarrow{\pa}_{D}
{\bar Z}^{B}  =
t^{-2} E^{AB},        
\eeq
which implies in turn,
\beq
\label{e3.21}
( \overrightarrow{\pa}_{A} \bar{Z} ^{C} ) E_{CD}
( \bar{Z}^{D}\overleftarrow{\pa}_{B}) = t^{-2} E_{AB}.    
\eeq Also, we have \beq \label{e3.22} D \bar{Z}^{A}  =  ( D Z^{B}  -
T \mathcal{ S } \overleftarrow{\pa}_{\theta} N
Z^{B} ) \overrightarrow{\pa}_{B} \bar{Z}^{A},      
\eeq
where we have used (\ref{e8}).
By inserting  (\ref{e3.22}) into (\ref{e3}), and using (\ref{e3.21}), we get
\beq
\label{e3.23}
D Z^{A}  +  t^{2} ( T \mathcal{ S }, Z^{A} )  - T \mathcal{ S }
\overleftarrow{\pa}_{\theta} N Z^{A}  =  0.       
\eeq
In terms of the extended antibracket  (\ref{AnBtau}), the equation (\ref{e3.23}) takes
immediately its desired canonical  form (\ref{e10}).

Although we have shown that all the motion equations generated by
the action (\ref{S})/(\ref{e1}) have the canonical form, that is not the case
for functional derivatives of the action taken at the general
trajectory in the phase space. Instead, the functional derivatives
do satisfy the so-called  canonical "Ward  Identities",
\beq
\label{theta}
\frac{\delta}{\delta \theta }\; \Sigma   = D \ln t  + ( T \mathcal{ S },  \ln t)_{\tau},
\eeq
\beq
\label{3.24}
\left[ t \frac{\delta}{ \delta t } + ( N Z^{A} )\frac{\delta}{\delta Z^{A}} \right] \Sigma
=  D \theta +
( T \mathcal{ S } , \theta )_{\tau},   
\eeq
\beq
\label{3.25}
\left[  t^{2} E^{AB}\frac{\delta}{\delta Z^{B}} + ( N Z^{A} ) \frac{\delta}{\delta
\theta} \right] \Sigma  =
 \left[ D Z^{A} + ( T \mathcal{ S },  Z^{A} )_{\tau} \right] (-1)^{ \varepsilon_{A} }.
\eeq
In the right-hand sides in these relations, one can recognize
the canonical form of the left-hand  sides  of the motion equations
(\ref{e7}), (\ref{e9}), (\ref{e10}). Now,  let  $[ F, G ]_{\tau}$  be the $\tau$
- extended  antibracket  in the space  of  functionals, generated
by  the ultralocal  $\tau$ - extended  antibracket $( F, G)_{\tau}$,
\beq
\nonumber
&&[ F, G ]_{\tau} =\int  [du] [d\xi] F
\left[\frac{\overleftarrow{\delta }}{ \delta Z^{A}}\; t^{2} E^{AB}
\frac{\overrightarrow{\delta }}{ \delta Z^{B}}\; +\right.\\
\label{3.26}
&&\left. +\frac{\overleftarrow{\delta }}{\delta Z^{A}} \;( N Z^{A} )
\frac{\overrightarrow{\delta }}{ \delta \theta}-
\frac{\overleftarrow{\delta }}{ \delta \theta} \;( N Z^{A} )
\frac{\overrightarrow{\delta }}{\delta Z^{A}}  +
\frac{\overleftarrow{\delta}}{\delta \ln t}\;\frac{\overrightarrow{\delta }}{ \delta \theta }
 - \frac{\overleftarrow{\delta }}{ \delta \theta}\; \frac{\overrightarrow{\delta }}
{\delta \ln t } \right]  G.     
\eeq
Due to the canonical "Ward Identities" (\ref{theta}), (\ref{3.24}), (\ref{3.25}), the action
$\Sigma$ does satisfy the functional
master equation in terms of the functional $\tau$ - antibracket
(\ref{3.26}),
\beq
\label{3.27}
\frac{1}{2} [ \Sigma, \Sigma ]_{\tau}  =  \int  [du]  [d\xi]
\left[ D \mathcal{L } +
\frac{1}{2} (T \mathcal{S},T \mathcal{S} )_{\tau} \right]   =   0.    
\eeq
In  terms  of  the  functional  antibracket   (\ref{3.26}), the
relations (\ref{theta}), (\ref{3.24}),  (\ref{3.25}) rewrite   as
\beq
\label{caneq}
[ \Sigma,  \Gamma  ]_{\tau}  =  \nabla  \Gamma,
\quad   \nabla   =   D  + {\rm ad}_{\tau} ( T \mathcal{ S } ), \quad
\Gamma  =  \{ \ln t,  \theta ,   Z^{A}\}.
\eeq
Due  to  the  Jacobi
identity  for  the  functional  antibracket, together  with  the
functional master  equation  (\ref{3.27}),  it follows from the
equations (\ref{caneq}) that  the compatibility  relations hold
\beq
[ \Sigma,  \nabla  \Gamma  ]_{\tau}   =   0.
\eeq
The first equality
in (\ref{3.27}) holds due to the following integral identity,
\beq
\label{3.31}
\int [du] [d\xi] (\nabla \Gamma^{\alpha}) (\nabla \Gamma^{\beta})\;
\omega_{\beta \alpha} \;(-1)^{\varepsilon_{\alpha}} = 0,
\eeq
where
\beq
\label{3.32}
\omega^{\alpha \beta} = ( \Gamma^{\alpha}, \Gamma^{\beta} )_{\tau},
\eeq
while $\omega_{\alpha \beta}$ is an inverse to (\ref{3.32}).

In  terms  of  the third  in  (\ref{caneq})  and  the metric  (\ref{3.32}),  the
functional  antibracket   (\ref{3.26})  rewrites   as
\beq
\label{3.33}
[ F, G ] _{ \tau }   =   \int  [ du ] [d \xi ]
F  \frac{\overleftarrow{\delta }}{\delta \Gamma^{ \alpha }}\;
\omega^{ \alpha \beta }\; \frac{\overrightarrow{\delta }}{\delta \Gamma^{ \beta }}  G   =
\int  [ du ] [d \xi ]  [ F,  \Gamma^{\alpha} ]_{ \tau }\;  \omega_{ \alpha
\beta }\;  [ \Gamma^{ \beta },  G ]_{ \tau }.     
\eeq
In  particular,  due  to  the  first  in  (\ref{caneq}),  we have  from  (\ref{3.33})  as
applied  for    $F  =  G  =  \Sigma$,
\beq
[ \Sigma,  \Sigma ]_{ \tau }  =  \int [ du ] [ d\xi ]
( \nabla \Gamma^{\alpha}  )  (\nabla \Gamma^{ \beta } ) \omega_{ \beta
\alpha } (-1)^{ \varepsilon_{ \alpha } }=\int [ du ] [ d\xi ]
[2D{\cal L}+(T{\cal S},T{\cal S})_{\tau}],   
\eeq
so  that  the  identity  (\ref{3.31})  does  imply  the  functional  master
equation  (\ref{3.27})  to  hold.

Notice that all the above reasoning  did use no further restrictions
to the Euler operator $N$. However, it follows from the second in (2.8), that
\beq
\label{e3.24}
N = N_{0}  +  2 {\rm ad}( F ) , \quad N_{0}  =  Z^{A} \pa_{A}, \quad   \varepsilon( F ) = 1,
\quad \pa_{A} \Delta F  =  0.      
\eeq
The most general form allowed for $F$ is
\beq
\label{e3.25}
 2 F  =  Z^{A} F_{AB} Z^{B}  +  \Delta Y ,\quad  \pa_{C} F_{AB} = 0,   
\eeq
\beq
\label{e3.26}
\varepsilon( F_{AB} )  = \varepsilon_{A}  +  \varepsilon_{B} + 1,\quad
\varepsilon( Y )  = 0,    
\eeq
\beq
\label{e3.27}
F_{AB} = F_{BA} (-1)^{ \varepsilon_{A} \varepsilon_{B} }.    
\eeq
As the operator $( N_{0} - 2 )$ does differentiate the antibracket, we have
\beq
\label{e3.28}
[ N_{0}, {\rm ad}( F ) ]  =  {\rm ad}( ( N_{0} - 2 ) F ).        
\eeq
In  the sense of  (\ref{e3.28}),  the  simplest  possibility in  (\ref{e3.25})  is
\beq
\label{e3.29}
 Y =  0, \quad  2 F = Z^{A} F_{AB} Z^{B},       
\eeq
which we will assume from now on.  It follows from (\ref{e3.28}), that, in the case
(\ref{e3.29}),
\beq
\label{e3.30}
[ N_{0}, {\rm ad}( F ) ] = 0,    
\eeq
which implies for the zero modes
\beq
\label{e3.31}
\bar{Z}^{A} = \exp\{ - ( \ln t ) 2 {\rm ad}( F ) \} \exp\{ - ( \ln t ) N_{0} \} Z^{A}  =
S^{A}_{\;\;B} t^{-1} Z^{B},      
\eeq where $S^{A}_{\;\;B}$ is an antisymplectic matrix given by
\beq
\label{e3.32} S^{A}_{\;\;B}  =  ( \exp\{ ( \ln t ) G \}
)^{A}_{\;\;B}, \quad
G^{A}_{\;\;B}  =  2 E^{AC}F_{CB},      
\eeq
so that we have,
\beq
\label{e3.33}
(-1)^{ ( \varepsilon_{A} + 1 ) \varepsilon_{C} } S^{A}_{\;\;C} E_{AB} S^{B}_{\;\;D}
= E_{CD}.     
\eeq

Notice  that,  in the  case  (\ref{e3.29}),  one can deduce  (\ref{e10})  from (\ref{e3})
even  more explicitly  by  making  use of  the representation  (\ref{e3.31})  for the  zero
modes.

In conclusion, we would like to clarify the following.  In the
present article, we did proceed with the action (\ref{S}) / (\ref{e1}) whose
kinetic part was defined in terms of the zero  - modes to the
extended Euler operator $N_{\tau}$.  It is a characteristic feature
of such an action that the original motion equations are non
certainly canonical in their form with  respect  to the extended
$\tau$ - antibracket.  The latter circumstance requires for special
technique as to transform the motion equations to their desired
explicit canonical form, as developed in Section 3.  The main
advantage of the approach based on the use of the zero-modes, is
that the antibracket  algebra of the zero modes is very  simple. On
the other hand, in terms of the variables, the third in (\ref{caneq}), one
can always use the action ( see also (\ref{3.32}) )
\beq
\label{LGamma}
\mathcal{ L }  =  \Gamma^{\alpha} \bar{ \omega }_{ \alpha \beta }
D \Gamma^{\beta} (-1)^{\varepsilon_{ \beta} }  +  T \mathcal{ S },
\eeq
\beq
\bar{\omega}_{ \alpha \beta } =
( \Gamma^{\gamma} \pa_{\gamma}  +  2 )^{-1} \omega_{\alpha \beta },
\eeq
\beq
\pa_{\gamma} \omega_{\alpha \beta} (-1)^{ \varepsilon_{ \gamma } \varepsilon_{\beta} }  +
{\rm cyclic \;\; perm.} ( \alpha, \beta, \gamma ) = 0.   
\eeq
With the use of  a rather complicated technique, the  action (\ref{LGamma})
yields the explicit canonical motion equations,
\beq
\nabla \Gamma^{\alpha}  =  0,     
\eeq
where $\nabla$  was defined in the second in  (\ref{caneq}). In
the action (\ref{LGamma}), the kinetic part involves a non-trivial
integral operator as  applied to $\omega_{ \alpha \beta }$. It seems
rather difficult to realize how the complicated structure of the
kinetic part in  the action (\ref{LGamma}) could be related
naturally to the geometric objects of the deformed  sigma model.  We
think, however, that the approach based on the use of the zero
modes, together with the approach based on the action
(\ref{LGamma}), do describe   complementary aspects of the whole
geometric construction.

\section*{Acknowledgments}
\noindent
I. A. Batalin would like  to thank Klaus Bering of Masaryk
University for interesting discussions. The work of I. A. Batalin is
supported in part by the RFBR grants 14-01-00489 and 14-02-01171.
The work of P. M. Lavrov is supported in part by the Presidential  grant 88.2014.2 for LRSS and by
the RFBR grant 15-02-03594.


\begin {thebibliography}{99}
\addtolength{\itemsep}{-8pt}

\bibitem{BV}
I. A. Batalin and G. A. Vilkovisky, {\it Gauge algebra and
quantization}, Phys. Lett. {\bf B102} (1981) 27.

\bibitem{BV1}
I. A. Batalin and G. A. Vilkovisky, {\it Quantization of gauge
theories with linearly dependent generators}, Phys. Rev. {\bf D28} (1983)
2567.

\bibitem{BT}
I. A. Batalin and I. V. Tyutin, {\it On local quantum deformation of antisymplectic
differential},
Int. J. Mod. Phys. {\bf A9} (1994) 517.

\bibitem{LSh}
D. A. Leites and I. M. Shchepochkina, {\it How to quantize the antibracket},
Theor. Math. Phys. {\bf 126} (2001) 281.

\bibitem{KoT}
S. E. Konstein and I. V. Tyutin, {\it
Deformations and central extensions of the antibracket superalgebra},
 J. Math. Phys. {\bf 49} (2008) 072103.

\bibitem{KoT1}
S. E. Konstein and I. V. Tyutin, {\it The deformations of nondegenerate
constant Poisson bracket with even and odd deformation parameters},
arXiv:1001.1776[math.QA]

\bibitem{KoT2}
S. E.  Konstein and I. V. Tyutin, {\it The deformations of antibracket
with even and odd deformation parameters, defined on the space $DE_{1}$},
arXiv:1112.1686[math-ph].

\bibitem{AKSZ}
 M.  Alexandrov,  M.  Kontsevich,  A  Schwarz,  and  O.  Zaboronsky,
 {\it The Geometry of the master equation and topological quantum field theory},
Int.  J.  Mod.  Phys. {\bf A12} (1997) 1405.



\bibitem{BB2}
I. A. Batalin and K. Bering,
{\it Path integral formulation with deformed antibracket},
Phys. Lett. {\bf B694} (2010) 158.


\bibitem{BL}
I. A. Batalin and P. M. Lavrov, {\it Does the nontrivially deformed field-antifield formalism
exist?}, Int. J. Mod. Phys. {\bf A30} (2015), arXiv:1502.07417[hep-th].

\bibitem{Witten}
E. Witten, {\it A note on the antibracket formalism}, Mod. Phys. Lett. {\bf A5}(1990) 487.

\bibitem{SS}
P. Schaller and T. Strobl, {\it Poisson structure induced (topological) field theories},
  Mod. Phys. Lett. {\bf A9} (1994) 3129.

\bibitem{CF}
A. S. Cattaneo and G. Felder, {\it A path integral approach to the Kontsevich
quantization formula}, Commun. Math. Phys. {\bf 212} (2000) 591.

\bibitem{LO}
A. M . Levin and M. A. Olshanetsky, {\it Hamiltonian algebroid symmetries in W-gravity and
Poisson sigma-model}, arXiv:hep-th/ 0010043.

\bibitem{BM4}
I. A. Batalin  and  R. Marnelius, {\it Generalized Poisson sigma models},
Phys. Lett.  {\bf B512} (2001) 225. 


\bibitem{BM5}
I. A. Batalin  and  R. Marnelius, {\it Superfield algorithms for
topological field theories}, Michael  Marinov memorial volume, M.
Olshanetsky, A. Vainstein [Eds.] WSPC (2002); [hep-th/0110140].

\bibitem{BLN}
L. Baulieu, A. S. Losev and N. A. Nekrasov, {\it Target space symmetries in topological
theories. I.}, JHEP {\bf 0202} (2002) 021.

\bibitem{BG}
G. Barnich and M. Grigoriev, {\it First order parent formulation
for generic gauge field theories},
JHEP {\bf 1101} (2011) 122.



\end{thebibliography}
\end{document}